\documentclass[preprint,aps,showkeys,preprintnumbers,onecolumn,showpacs,amsmath,amssymb,floatfix,endfloats*]{revtex4}

\usepackage{graphicx}
\usepackage{dcolumn}
\usepackage{bm}

\begin{document}


\title{Cavity-resonant excitation for efficient single photon generation}

\author{M. Kaniber} \email{kaniber@wsi.tum.de} \homepage{http://www.wsi.tum.de}
\author{A. Neumann}
\author{A. Laucht}
\author{M. Bichler}
\author{M.-C. Amann}
\author{J. J. Finley} 
\affiliation{Walter Schottky Institut and Physik Department, Technische Universit\"at M\"unchen, Am Coulombwall 3, D-85748 Garching, Germany}
\date{\today}

%
\begin{abstract}
We present an efficiently pumped single photon source based on single quantum dots (QD) embedded in photonic crystal nanocavities. 
Resonant excitation of a QD via a higher order cavity mode results in a 100$\times$ reduced optical power at the saturation onset of the photoluminescence, compared with excitation at the same frequency, after the cavity mode is detuned. Furthermore, we demonstrate that this excitation scheme leads to selective excitation of QDs coupled to the cavity by comparing photoluminescence and auto-correlation spectra for the same excitation wavelength with and without the cavity mode. This provides much cleaner conditions for single photon generation. 
\end{abstract}

\maketitle
%
%
Investigations of single quantum dots coupled to high quality, low mode volume solid-state nanocavities attract much interest in current research, since they offer a promising route towards highly efficient single photon turnstile devices \cite{Michler00}, for a wide range of applications in quantum information science \cite{Knill01, Gisin02}. However, the required spectral \cite{Hofbauer07} and spatial \cite{Badolato05} coupling of a single quantum emitter to the cavity necessitate demanding nano-fabrication of the cavities as well as an in-situ spectral tuning technique. Nevertheless, recent results prove that single quantum dots inside a photonic bandgap material give rise to highly efficient single photon emission \cite{Kaniber08a}, without the need for sophisticated nanofabrication techniques. For more demanding quantum optical applications \cite{Santori02} which require single quantum dot (QD)-cavity coupling, low QD densities make it statistically very unlikely to obtain a dot spatially coupled to the cavity mode without the need for deterministic positioning of individual dots. In contrast, higher QD densities, which result in higher probability for single dot-cavity coupling, would give rise to more complicated emission spectra and thus degrade the purity of single photon state preparation. Therefore, cavity-resonant excitation could be exploited for realizing \textit{selective} and \textit{efficient} excitation of a single QD inside a nanocavity as proposed recently by Nomura et al \cite{Nomura06}.\\ 
%
%
In this letter, we demonstrate an optical excitation scheme for realizing an efficiently pumped single photon source based on single QDs coupled to a two-dimensional photonic crystal (2D-PC) nanocavity. Comparative micro-photoluminescence ($\mu$-PL) measurements of the same QD transition subject to quasi-resonant excitation in the wetting layer (WL) and cavity-resonant excitation close to a higher order cavity mode show a striking reduction of the saturation onset for the QD emission by a factor of $100\times$. Furthermore, we evidence that such cavity-resonant excitation \textit{selectively} excites quantum dots which are spatially strongly coupled to the cavity field leading to a reduction of the number of emission peaks in the PL spectra and, therefore, also to a suppression of background emission which limits the single photon state purity.\\
%
%
The sample studied is grown by molecular beam epitaxy and consists of the following layers grown on a semi-insulating GaAs wafer:  an undoped GaAs buffer is followed by a $500~nm$ thick Al$_{0.8}$Ga$_{0.2}$As sacrificial layer and a $180~nm$ thick GaAs waveguide, at the midpoint of which a single layer of self-assembled In$_{0.5}$Ga$_{0.5}$As quantum dots was incorporated. A 2D-PC was formed by patterning a triangular array of cylindrical air holes using electron-beam lithography and reactive ion etching. The lattice constant of the PC is $a=280~nm$ and the air hole radius is $r=0.32a$.  Nanocavities were established by introducing three missing holes to form an $L3$ cavity \cite{Akahane03}. Finally, free standing GaAs membranes were formed by an HF wet etching step. An overview of the different fabrication steps can be found in ref. \cite{Kress05}.\\
%
%
The sample was mounted in a liquid He-flow cryostat ($T=15~K$) and excited by $2~ps$ duration pulses delivered by a Ti:Sapphire laser with a repetition rate of 80 MHz. The QD $\mu$-PL  was collected via a $100\times$ microscope objective ($NA=0.8$) providing a spatial resolution of $\sim700~nm$, spectrally analyzed by a $0.55~m$ imaging monochromator and detected with a Si-based, liquid nitrogen ($LN_2$) cooled CCD detector. A pair of similar detectors in Hanbury Brown and Twiss configurations were used for photon auto-correlation measurements.\\
%
\begin{figure}
\includegraphics[width=\columnwidth]{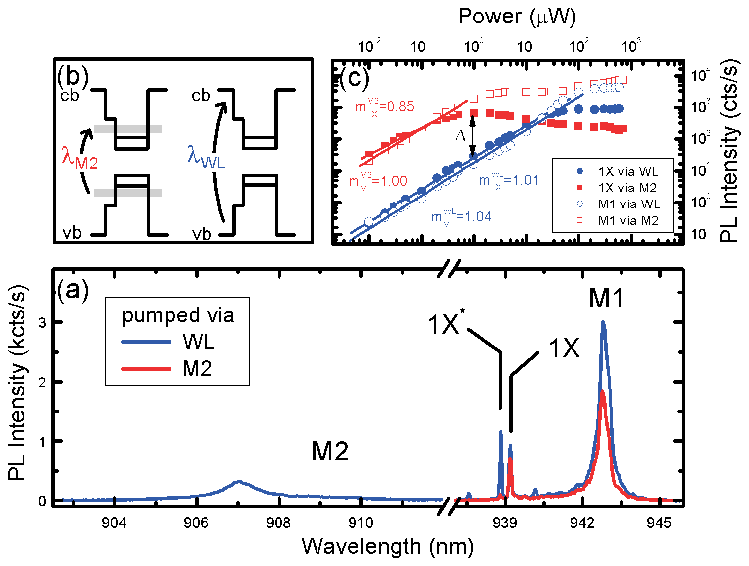}
\caption {(Color online) (a) compares $\mu$-PL spectra of the same QD with quasi-resonant excitation into the WL  
          (blue) and with cavity-resonant excitation via M2 (red). (b) Schematic illustrations of cavity-resonant and quasi-resonant
          excitation. (c) shows the $\mu$-PL intensity as a function of excitation power for $1X$ (full symbols) and $M1$ (open symbols) 
          with cavity-resonant excitation (red) and quasi-resonant excitation (blue).}
\end{figure}
%
%
In Fig. 1, we compare $\mu$-PL spectra recorded using quasi-resonant excitation into the two-dimensional WL (b - right panel) and cavity-resonant excitation via a higher energy mode (b - left panel). For quasi-resonant excitation (Fig. 1a - blue), the spectrum exhibits two QD transitions labeled $1X$ ($\lambda_{1X}=939.18~nm$) and $1X^*$ ($\lambda_{1X^*}=938.82~nm$) besides a couple of very weak emission features. We attribute their transitions to different single exciton configuration of the same QD. The much broader peaks labeled $M1$ ($\lambda_{M1}=942.78~nm$, $Q_{M1}^{PL}=2000$) and $M2$ ($\lambda_{M2}=907.01~nm$, $Q_{M2}^{PL}=900$) arise from different cavity modes of the investigated $L3$ cavity. By applying cavity-resonant excitation via $M2$ (Fig. 1a - red), the transition $1X^*$ and the weak emission features become strongly suppressed, whilst $1X$ and $M1$ dominate the spectrum. The suppression of $1X^*$ with cavity-resonant excitation and the blue shift ($E_{1X^*-1X}=0.5~meV$) compared to $1X$ indicate that this transition is likely to arise from a positively charged exciton \cite{Finley01}. In Fig. 1c, we plot the PL intensity of $1X$ and $M1$ (on a double logarithmic scale) as a function of the excitation power for both WL and cavity-resonant excitation. With quasi-resonant excitation (full circles) we observe a linear dependence of $1X$ ($m_{1X}^{WL}=1.01$) with a saturation power $P_{sat}=100~\mu W$. This means that the probability per laser pulse to create a single electron-hole pair in the QD is larger than one. In strong contrast, with excitation via the cavity mode $M2$ (full squares), $1X$ shows a sub-linear behavior ($m_{1X}^{M2}=0.85$) below saturation and a strongly reduced saturation power of $P_{sat}=1~\mu W$. The reduction of the saturation power by two orders of magnitude means that we pump excitation into the dot with a $100\times$ higher efficiency. In addition to the higher saturation level for cavity-resonant excitation, the PL intensity with cavity-resonant excitation at $1~\mu W$ is a factor $\Delta=25$ higher than for quasi-resonant excitation. This shows clearly the potential for realizing an efficiently pumped single photon source by exploiting such cavity-resonant excitation.\\
The decrease of the PL intensity above saturation is indicative for the existence of a non-resonant coupling mechanism between the QD transition $1X$ and the cavity mode $M1$ as recently published in ref. \cite{Hennessy07, Press07, Kaniber08b}. This behavior indicates that a mechanism exists whereby feeding of the mode by the QD emission takes place even though the dot is spectrally detuned from the cavity mode. The presence of this feeding process can be directly observed in the power dependence of the cavity mode $M1$ recorded using cavity-resonant excitation via $M2$ (open squares) in Fig. 1c. Clearly, the intensity in the saturation regime increases whilst the QD transition reduces. The fact that this process is not observed for quasi-resonant excitation (open circles in Fig. 1c) is a strong indication that a final state continuum must be involved in this feeding mechanism (see ref. \cite{Kaniber08b}). Such continuum states are much more easily filled via cavity-resonant excitation than via quasi-resonant excitation into the two-dimensional WL states.\\ 
%
%
\begin{figure}
\includegraphics[width=\columnwidth]{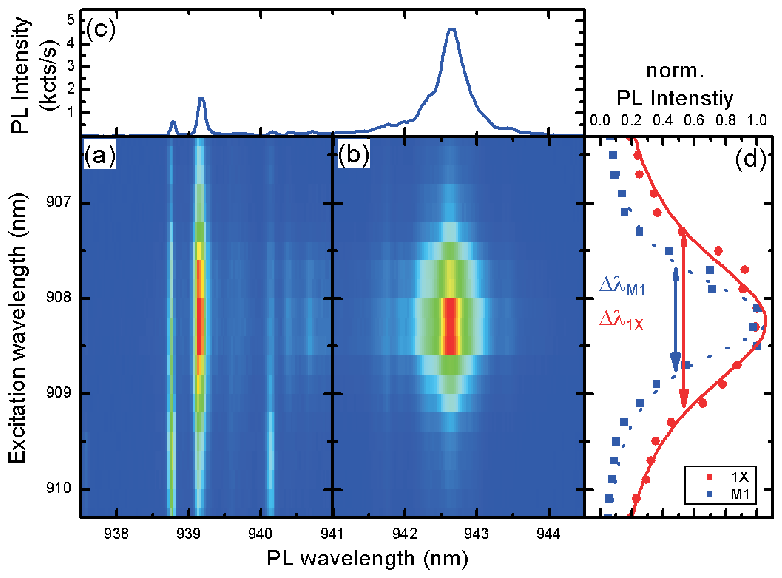}
\caption {(Color online) (a) and (b) show MPLE spectra of $1X$ and $M1$, respectively. (c)            
          $\mu$-PL spectrum of $1X$ and $M2$ at resonance ($\lambda_{res}=908.25~nm$). (d) PLE spectra of $1X$ 
          (circles) and $M1$ (squares).} 
\end{figure}
%
%
Resonant excitation via a higher order cavity mode enhances the absorption of the incident light at the high electric field regions inside the defect cavity \cite{Nomura06, Oulton07}. This enhancement effectively reduces the excited region of the sample due to concentration of the excitation power inside the cavity region. Therefore, quasi-resonant excitation via a higher energy cavity mode enables us to selectively excite quantum dots that are spatially coupled to an electric field maximum of that specific mode. To check this expectation, we performed multichannel photoluminescence excitation (MPLE) measurements in which we scanned the excitation wavelength through the higher order cavity mode $M2$ from $\lambda_{exc}=906.3~nm$ to $910.3~nm$ and simultaneously recorded the PL spectra around the lower energy mode $M1$. The result is shown in Figs. 2a and 2b, where the detection and excitation wavelengths are plotted on the x- and y-axis, respectively, and the PL intensity is encoded in false color. We observe a common maximum in the intensity of both $1X$ and $M1$ when exciting spectrally in resonance with the higher energy mode $M2$ at $\lambda_{exc}=\lambda_{M2}=908.25~nm$ \cite{comment01}. This observation indicates that the QD transition is spatially well coupled to the cavity field maximum of the mode $M2$. Moreover, it demonstrates that the excitation pumped into the cavity at $\lambda_{M2}$ is directly transferred to the QD transition. For comparison, in panel (c) we plot a PL spectrum recorded at $\lambda_{exc}=\lambda_{M2}$. In Fig. 2d, we present the normalized PL intensity of $1X$ (circles) and $M1$ (squares) as a function of the excitation wavelength. Both curves should, in principle, reflect the spectral form of cavity mode $M2$ via which the system has been excited. By fitting the data points for the lower energy mode $M1$, we obtain a quality factor of $Q_{M2}^{via~M1}=850$, which is in very good agreement with the value directly obtained for $M2$ in the PL studies discussed in Fig. 1. However, the quality factor obtained via the QD transition $1X$ ($Q_{M2}^{via~1X}=470$) is much lower than the one from the PL measurements since this measurement has been conducted with an excitation power in the saturation regime of $1X$, where the PL intensity decreases as shown in Fig. 1c.\\ 
%
%
\begin{figure}
\includegraphics[width=\columnwidth]{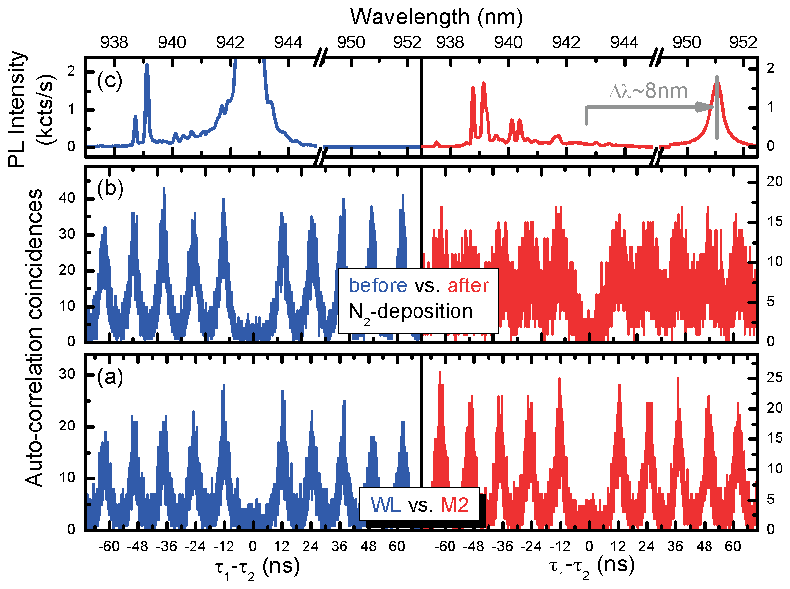}
\caption {(Color online) (a) compares auto-correlation spectra recorded on $1X$ with quasi-resonant excitation (left panel) and cavity-resonant  
          excitation (right panel). (b) compares auto-correlation spectra recorded on $1X$ before (left panel) and after (right panel) $N_2$-deposition 
          for constant excitation wavelength. (c) compares $\mu$-PL spectra before (left panel) and after (right panel) $N_2$-deposition.}
\end{figure}
%
%
To test the feasibility of the cavity-resonant excitation scheme for realizing an efficiently pumped single photon source, we performed photon auto-correlation spectroscopy for probing the photon statistics of the $1X$ transition. In Fig. 3a, we compare auto-correlation measurements recorded from $1X$ with quasi-resonant excitation into the WL (left panel) and cavity-resonant excitation via $M2$ (right panel). For both excitation schemes we chose optical powers such that the excitation level of the dot was equivalent, near the PL saturation level. By reference to Fig. 1c, this is $100\mu W$ and $1\mu W$ for quasi-resonant and cavity-resonant excitation, respectively. In both cases we recorded auto-correlation traces for an integration time $t=5000~s$ and obtained reasonably fair suppression of the multi-photon emission probability of $\sim25\%$ for both quasi-resonant and cavity-resonant excitation. Although we observe qualitatively similar results for both excitation schemes, it is important to note that we reduced the excitation power by a factor of $100\times$ when exciting via the higher order mode $M2$, when compared to exciting into the WL. This clearly demonstrates the advantage of cavity mode pumping for optimizing the ratio of PL saturation intensity to incident optical power.\\
The comparison of two excitation schemes with completely different spectral positions and as a consequence different optical absorption coefficients does not unambiguously prove the concept of cavity-resonant pumping since a local maximum of the absorption coefficient (e.g. due to excited QD states) at $\lambda_{M2}$ would also explain the pronounced decrease in saturation power. However, a controlled thin layer deposition of molecular $N_2$ onto the sample surface \cite{Mosor05}, enables us to deterministically detune the cavity modes from their original spectral positions by adsorption into the PC. This allows us to test if the reduction in excitation power  needed to generate clean single photons is really caused by the existence of a cavity mode ($M2$) or only due to the difference in optical absorption coefficients for both excitation schemes. Thus, we performed auto-correlation measurements on $1X$ for exactly the same excitation conditions ($\lambda_{exc}=\lambda_{M2}$) \textit{before} and \textit{after} the shift of $M2$. The results presented in Fig. 3b are for excitation before (left panel) and after (right panel) the $N_2$-deposition and same integration times $t=7200~s$. We observe an enhanced background between two adjacent peaks in the auto-correlation spectrum after the $N_2$-deposition (b - right). This is due to the longer lifetime of the $1X$ transition after the mode $M1$ is also detuned from $1X$. Beside this, we observe again qualitatively very similar results for both measurements. This measurement confirms that the reduction of the excitation power by two orders of magnitude is, indeed, related to the existence of the cavity mode at $\lambda_{M2}$ since the onset of saturation was found to shift from $1~\mu W$ (before $N_2$-deposition) to $100~\mu W$ (after $N_2$-deposition) (data not shown here). The corresponding PL spectra presented in Fig. 3c, clarify that $M2$ has indeed shifted by $\sim8~nm$ towards longer wavelength after the injection of $N_2$. Clearly, the PL spectrum after the deposition becomes much more complicated, indicating that the excitation is \textit{not} as selectively directed towards a single QD state as it has been before the deposition. Thus, cavity-resonant pumping suppresses very effectively emission of quantum dots which are spatially less coupled to the cavity and, therefore, leads to clear \textit{selective} excitation of dots spatially coupled to the electric field maximum of the cavity mode.\\
%
%
In summary, we have demonstrated the practicability of a cavity-resonant excitation scheme that leads to efficiently pumped single photon generation based on quantum dots coupled to PC nanocavities. The excitation via a higher order cavity mode results in comparable PL intensities of the QD and multiphoton emission probabilities, as compared to non-resonant excitation, whilst the necessary optical power needed to generate the same excitation level in the dot is reduced by a factor of up to $100\times$. Furthermore, we showed that this new excitation scheme leads to suppression of the emission from quantum dots which are spatially decoupled from the electric field inside the cavity allowing for much more direct excitation of a specific QD state. In addition to efficient single photon emission in these kinds of systems \cite{Kaniber08a}, we believe that this technique may also enables new perspectives on the route towards an efficient optically pumped single photon source, since it provides a method for clean and efficient quantum state preparation.\\
%
%
We acknowledge financial support of the Deutsche Forschungsgemeinschaft via the Sonderforschungsbereich 631, Teilprojekt B3 and the German Excellence Initiative via the ``Nanosystems Initiative Munich (NIM)".\\
%
%

%
%

\begin{thebibliography}{999}
\bibitem{comment01} The slight offset between of $\lambda_{M2}$ in PLE measurements compared to PL measurements is due an offset of the excitation laser frequency.
\end{thebibliography}
\end{document}